\documentclass[conference]{IEEEtran}
\IEEEoverridecommandlockouts
\usepackage{cite}
\usepackage{amsmath,amssymb,amsfonts}
\usepackage{algorithmic}
\usepackage{graphicx}
\usepackage{textcomp}
\usepackage{xcolor}
\usepackage{float}
\usepackage{multirow}
\usepackage[hidelinks]{hyperref}
\def\BibTeX{{\rm B\kern-.05em{\sc i\kern-.025em b}\kern-.08em
    T\kern-.1667em\lower.7ex\hbox{E}\kern-.125emX}}

\begin{document}

% \title{Employing Robust Statistics for Continuous Inertia Estimation from Ambient Synchrophasor Data\\
\title{A Practical Approach Towards Inertia Estimation Using Ambient Synchrophasor Data\\
% \title{Inertia Estimation from Ambient Synchrophasor Data in the Presence of Bounded Uncertainties\\
\thanks{This work was supported in part by the U.S. Department of Energy (DOE) grant 
% having
% % Energy’s Office of Energy Efficiency and Renewable Energy (EERE) under the Solar Energy Technology Office (SETO) 
% award number 
DE-EE0011372. 
The views expressed herein do not necessarily represent the views of the U.S. DOE or the U.S. Government.}
}

\author{
    \IEEEauthorblockN{Anushka Sharma\textsuperscript{1}, Anamitra Pal\textsuperscript{1}, Rajasekhar Anguluri\textsuperscript{2}, and Tamojit Chakraborty\textsuperscript{3}}
    \IEEEauthorblockA{\textsuperscript{1}\text{School of Electrical, Computer, and Energy Engineering}, \textit{Arizona State University}, {Tempe, AZ}}
\IEEEauthorblockA{\textsuperscript{2}\text{Department of Computer Science and Electrical Engineering}, \textit{University of Maryland, Baltimore County}, Baltimore, MD}
\IEEEauthorblockA{\textsuperscript{3}\textit{Siemens Energy}, Manchester, England, United Kingdom\\
    ashar561@asu.edu, anamitra.pal@asu.edu, rajangul@umbc.edu, tamojit91@gmail.com}
}

\maketitle

\begin{abstract}
Real-time tracking of inertia is important because it reflects the power system’s ability to withstand contingencies 
% which is crucial for maintaining frequency security.
and maintain frequency security.
% With the increasing number of converter-interfaced generation (CIG), tracking of inertia is essential for reliability of power systems. 
This paper proposes a practical approach to 
% effectively 
estimate inertia using ambient phasor measurement unit (PMU) data and a partitioned form of the swing equation.
% , relaxing the common assumption of constant mechanical power input. 
The approach accounts for (bounded) uncertainties in network parameters and PMU measurements, enabling precise estimation of inertia and damping constants, as well as mechanical power inputs. 
Instead of assuming constant mechanical power input throughout, the approach leverages 
% we leverage our 
knowledge of power system operations 
to determine intervals when it is \textit{actually} constant to maintain estimation consistency.
Simulation results on the IEEE 14-bus system and IEEE 39-bus system integrated with renewable energy sources affirm the method's accuracy and applicability.
\end{abstract}
%serves as a linchpin
\begin{IEEEkeywords}
Ambient data, Bounded data uncertainty, Frequency divider, Inertia estimation, Phasor measurement unit
\end{IEEEkeywords}

\section{Introduction}

% With the gradual replacement of centrally-located synchronous generation (SG) with spatially-distributed converter-interfaced generation (CIG), 
With spatially-dispersed
% spread 
converter-interfaced generation (CIG) gradually replacing centrally-located synchronous generation, it has become necessary to track system inertia in real-time \cite{article}. 
% \rajmargin{Ours is also model-based right?}
% This is because system inertia reflects the power system’s ability to withstand contingencies and is crucial for maintaining frequency security \cite{article}.
% \textit{Offline} methods for inertia estimation primarily comprise model-based approaches. However, they have been known to suffer from inaccuracies in the models (particularly, the black-box nature of CIGs) and/or uncertainties in the network parameters \cite{https://doi.org/10.1049/esi2.12122}. 
Real-time methods for inertia estimation primarily comprise phasor measurement unit (PMU) data-driven approaches, and can be 
% grouped under three categories: \textit{disturbance-based}, \textit{perturbation-based}, and \textit{ambient data-based} \cite{article2}. 
disturbance-based, perturbation-based, or ambient data-based \cite{article2}.
% Due to their dependence on events, a disturbance-based approach cannot be employed for continuously tracking system inertia. Similarly, performance of perturbation-based approaches is dependent on the judicious selection of the probing signal \cite{10483553},
% % (the signal should be enough to excite the system without causing any harm to the operation of the system) 
% % making them harder to apply to an actual power system. 
% limiting their practical use.
The dependence on events and judicious selection of the probing signal limit the practical use of the disturbance-based and perturbation-based approaches, respectively \cite{10483553}.
% Due to their dependence on events, a disturbance-based approach cannot be employed for continuously tracking system inertia, while performance of perturbation-based approaches is dependent on the judicious selection of the probing signal,
% % (the signal should be enough to excite the system without causing any harm to the operation of the system) 
% % making them harder to apply to an actual power system. 
% limiting their practical use \cite{10483553}.
The ambient-data based inertia estimation approaches often assume constant mechanical power input 
%($p_{m}$) 
%to be a constant 
\cite{9760051} and ignore uncertainties in the network parameters.
% , which is not the case in reality.
This paper advances the state-of-the-art by performing
% The
% % Considering all of these facets of the problem, the 
% aim of this paper is to perform 
\textit{real-time inertia estimation using ambient PMU data while accounting for the variations in mechanical power inputs and uncertainties in network parameters as well as synchrophasor measurements.}
We first estimate rotor speeds and acceleration from PMUs 
% placed at very few
% a few of the 
% network buses by leveraging
% by extending 
by extending
the frequency divider formula (FDF) \cite{7470466}, and then solve the resulting linear estimation problem in the presence of bounded uncertainties. 
Next, we determine the
% The first stage determines 
inertia constant ($H$) and damping constant ($D$) of individual generators from the swing equation 
% by accounting for uncertainties in the network parameters 
while \textit{treating} the mechanical power input ($p_{m}$) as a constant.
However, since we do not know beforehand the length of the actual windows for which $p_{m}$ is constant, considerable fluctuations occur in the $H$ and $D$ estimates. 
% obtained in this stage.
We use the fluctuations in the estimates
% In the second stage, these variations in $H$ and $D$ estimates 
% % obtained from the first stage 
% are used 
to figure out time intervals for which $p_{m}$ \textit{is} an unknown constant, and then solve for all three unknown parameters ($H$, $D$, $p_m$) through a \textit{partitioned form} of the
% modified version of the 
swing equation. 
We draw insights from actual power system operations to ensure practicality of the proposed approach. 
% \par This paper proposes an approach to estimate inertia using the frequency spatial correlation incorporating bounded data uncertainties and a modified swing equation to account for the changes in mechanical power. 
Our main contributions are:
% \begin{enumerate}
%     \item \textit{Damping}: Ensuring grid stability in scenarios with high penetration of inverter-based resources (IBRs) requires real-time management of both inertia and damping \cite{10477536}. Our approach goes beyond traditional inertia estimation by including the calculation of the damping constant, enhancing the assessment of system stability.
%     \item \textit{Mechanical Power}: Unlike previous methods that assume mechanical power remains constant, we recognize that this is often not the case in practical applications. Our approach accounts for variations in mechanical power, demonstrating that changes in this power can be inferred from fluctuations in inertia and damping.
%     \item \textit{Bounded Data Uncertainties}: While uncertainties in frequency, network parameters, and load changes are inherent, they typically remain within specific bounds under standard operating conditions. We incorporate these bounded uncertainties into our framework, introducing a refined approach for accurate inertia and damping estimation.
% \end{enumerate}
\begin{enumerate}
    \item Along with inertia, we also determine damping constant and mechanical power input for the windows for which the mechanical power input is actually constant.
    \item We demonstrate that variations in mechanical power inputs can be inferred from  fluctuations in inertia and damping estimates leading to a relaxation of the assumption of constant mechanical power input throughout.
    \item We exploit the bounded nature of the perturbations in network parameters and PMU data to treat the estimation problems as bounded data uncertainty (BDU) problems. 
    % To the best of our knowledge, this is the first paper that solves the inertia estimation problem in this way. 
    % \rajmargin{The last line of contributions seems redundant, right. If someone else solved it; why would we do it again?}  
\end{enumerate}

\section{Proposed Approach for Rotor Speed and Acceleration Estimation}
% \section{Problem Setup and Preliminaries}
\label{sec 2}
% Introduced in \cite{7470466}, the
The FDF 
% introduced in \cite{7470466} 
establishes a linear relationship
between rotor speeds of synchronous machines and
% the 
frequency of network buses. Combining the FDF
% this formula 
with the knowledge of network topology and bus frequencies 
obtained via PMUs, 
% or simulation tools such as PSS/E) 
one can estimate the rotor speeds and acceleration,
% (and subsequently, acceleration), 
which can then be used to estimate inertia from the swing equation. 
% These speeds are then used to do inertia estimation, which we discuss in the next section.
However, network parameters are subject to changes and PMU measurements have errors, implying that the estimation problem will be subject to uncertainties.
These aspects of the problem are discussed in this section, while the proposed approach for inertia estimation is described in the next section.

\subsection{Frequency Divider and Spatial Correlation}
\label{FreqDC}
%The internal electromotive force (EMF) frequency or the rotor speed is used to estimate the bus frequency. The internal EMF is formed when the magnetic field of the rotor winding cuts that of the stator winding.
Consider a power system having $N$ buses and $M$ point of interconnection (POI) buses to which generators are connected.
% The network bus current injection and voltage are expressed by $I_B$ and $V_B$, generator current injection and EMF are expressed by $I_G$ and $E_G$. The admittance matrix ($\bold{Y}$) includes the submatrices in relation to transient reactance, line parameters, and system topology. 
% \begin{align}
% \begin{bmatrix}
%     I_G \\
%     I_B
% \end{bmatrix} = \label{nw_fd}
% \underbrace{\begin{bmatrix}
%     Y_{GG} & Y_{GB}\\
%     Y_{BG} & Y_{BB}
% \end{bmatrix}}_{\bold{Y}}
% \begin{bmatrix}
%     E_G \\
%     V_B
% \end{bmatrix}
% \end{align}
For such a system, the FDF can be mathematically expressed in the following way \cite{7470466}:
% s derived by taking the derivative of the second row of \eqref{nw_fd} and $|E_G| = |V_B| = 1$ p.u.,
\begin{align}
    0_{N} = B_{BG} (\omega_G(t)-\mathbf{1}) + B_{BB} (\omega_B(t)-\mathbf{1}), \label{fd}
\end{align}
   where $\omega_B(t)\in \mathbb{R}^N$ is the vector of bus frequencies; $\omega_G(t)\in \mathbb{R}^M$ is the vector of rotor speeds; $\mathbf{1}$ is an appropriately sized vector indicating nominal frequency (angular/bus) in p.u.; and the matrices are: 
    \begin{itemize}
        \item $B_{BG}\in \mathbb{R}^{N\times M}$ denotes the susceptance between the generators and the buses to which they are connected,
        \item $B_{BB}=B_{BUS}+B_{GG} \in \mathbb{R}^{N\times N}$, 
        \item $B_{BUS}\in \mathbb{R}^{N\times N}$ is the imaginary part of the complex-valued network admittance matrix,  
        \item $B_{GG}\in \mathbb{R}^{N\times N}$ is a diagonal matrix with $B_{GG}(i,i)\ne 0$ if a generator is present at bus $i$, and its value is the inverse of the generator's internal reactance.
    \end{itemize}

For ease of notation, we drop the time argument $t$ from both $\omega_G(t)$ and $\omega_B(t)$.
% $ w_G$ denotes rotor speed ($\mathbb{R}^{M \times 1}$)
% $ w_B$ denotes bus frequency ($\mathbb{R}^{N \times 1}$)
% $B_{BG} \in \mathbb{R}^{N \times M}$ representing susceptances between the generators and the buses \\
% $B_{BB} = B_{BUS} + B_{GG} (\in \mathbb{R}^{N \times N})$ \\
% $B_{BUS} \in \mathbb{R}^{N \times N}$, is the imaginary part of the network admittance \\
% $B_{GG} \in \mathbb{R}^{N \times N}$, is a diagonal matrix with $B_{GG}(i,i) \neq 0$, if a synchronous machine is at bus \textit{i}, its value is the inverse of its internal reactance. \\
Equation \eqref{fd} can now be partitioned as:
\begin{align}
\begin{bmatrix}
    0_{M} \\
    0_{N}
\end{bmatrix} = \label{nw_fd}
\begin{bmatrix}
    \operatorname{diag}(b_{Gb}) & B_{Bb}\\
    0_{N \times M} & B_{Bl}
\end{bmatrix}
\begin{bmatrix}
    \omega_G -\mathbf{1} \\
    \omega_B - \mathbf{1}
\end{bmatrix} 
\end{align}
where $\operatorname{diag}(b_{Gb}) \in \mathbb{R}^{M\times M}$ is a
% an $M\times M$-dimensional 
diagonal matrix whose 
% diagonal 
entries are the inverse of the 
% generators' internal reactances,
combined reactances of the generator's internal reactance and the reactance of the line and transformer located between the generator and the POI,
% are $1/({x_{dq}^{'}})$, (${x_{dq}^{'}}$ is the generator's internal reactance),
$B_{Bb}$ captures 
% is 
the relation between frequency of POI buses and rotor speeds, and $B_{Bl}$ relates the frequency of one bus with that of electrically connected buses.
% \rajmargin{check the paragraph below (2)}
% and $x_L+x_T$ is the combined reactance of the transmission line and transformer located between the generator and the POI. $diag(b_{Gb})$ contains connection between internal EMF and its POI, including the aggregate of transient reactance, reactances of generator step-up transformer (GSU) and transmission line ($b_{Gb} = 1/({x_{dq}^{'}} + x_L + x_T)$).
% \subsubsection{Relationship between rotor speed and measured bus frequency} 
From the first row of \eqref{nw_fd}, 
% and considering $\textit{i}^{th}$ generator and bus frequencies, 
we get:
% \begin{align}
%     \hat{w}_{G,i} = - b_{Gb,i}^{-1} B_{Bb,i} {w}_{B} \label{fd2}
% \end{align}
\begin{align}
    ({\omega}_{G} - \mathbf{1}) = - \operatorname{diag}(b_{Gb})^{-1} B_{Bb} ({\omega}_{B} - \mathbf{1}) \label{fd2}
\end{align}

%Now, 
If PMUs measure $L$ ($L\leq N$) of the $N$ bus frequencies,
% a sub-set of the bus frequencies, 
then exploiting the frequency spatial correlations that exist in the power system, we can write:
% the following equation can be written:
% where ${w}_{B}$ represents bus frequencies corresponding to POI.
% \begin{align}
%     w_{B, me}^{(i)} = E {w}_B \label{E_relation}
% \end{align}
\begin{align}
    \omega_{B, me} = E ({\omega}_B-\mathbf{1}) \label{E_relation}
\end{align}
where 
% In \eqref{E_relation}, 
$\omega_{B, me} \in \mathbb{R}^{L \times 1}$
% $w_{B, me}^{(i)} \in \mathbb{R}^{L \times 1}$ 
denotes the measured bus frequency deviations, and $E$ $\in \mathbb{R}^{L \times N}$ is 
% a sparse matrix whose elements are 0 or 1.
an identity-like matrix whose rows corresponding to bus frequencies that are not measured by PMUs, have been removed.
% , and $L$ corresponds to buses where PMUs are located.
% The correlation between rotor speed and measured network bus frequency is represented by,
Next, by combining \eqref{E_relation} with 
the second row of 
\eqref{nw_fd}, the following relation between
% $i^{th}$ 
the generators' rotor speed deviations and the measured bus frequency deviations can be established \cite{9851923}:
% \begin{align} 
%     \hat{w}_{G,i} = -{b_{Gb,i}}^{-1} B_{Bb,i} \begin{bmatrix}
% -E \\
% B_{Bl}
% \end{bmatrix}^+
% \begin{bmatrix}
% I_L \\
% 0_{N \times L}
% \end{bmatrix} w^{(i)}_{B,me}
% \end{align}
\begin{align}
    ({\omega}_G-\mathbf{1}) &= \left \{ \operatorname{diag}(b_{Gb})^{-1} B_{Bb} 
\begin{bmatrix}
    E \\
    B_{Bl}
\end{bmatrix}^+ 
\begin{bmatrix}
    I_L \\
    0_{N \times L} 
\end{bmatrix} \right \} \omega_{B,me}\label{rot_bus_freq}
\end{align}
where, $I_L$ $\in \mathbb{R}^{L \times L}$ is an identity matrix, and $[\cdot]^+$ denotes pseudo-inverse.
Finally, by denoting the terms in curly brackets in \eqref{rot_bus_freq} by $C\in\mathbb{R}^{M \times L}$, we  establish the following 
% linear 
relation between rotor speed and measured bus frequency deviations:
\begin{align}
    ({\omega}_G - \mathbf{1})  = C\omega_{B,me}
    \label{final_wg}
\end{align}
% % \subsubsection{Rotor speed estimation}
% Rotor speed of all generators is calculated by,
% \begin{align}
%     \hat{w}_G &= diag(b_{Gb})^{-1} B_{Bb}  
% \begin{bmatrix}
%     E \\
%     B_{Bl}
% \end{bmatrix}^+ 
% \begin{bmatrix}
%     I_L \\
%     0_{N \times L} 
% \end{bmatrix} w_{B,me}\\ \label{rot_bus_freq}
%     \hat{w}_G  &= A\: w_{B,me}
% \end{align}
% where, $A \in \mathbb{R}^{M \times M}$ shows a linear correlation between rotor speed and POI bus frequencies.

%Note that 
To uniquely estimate all rotor speeds using \eqref{final_wg}, at a minimum, every POI bus frequency must be measured by PMUs.
Going forward, we assume that is the case and set $L=M$.
Note that one can show $C$ is full rank in this case.
%is the case by setting $L=M$.
Finally, 
% since PMUs give rate-of-change-of-frequency measurements and $C$ is time-invariant (as long as topology does not change),
% (as long as the topology does not change, which is true for ambient data), 
\eqref{final_wg} can be used
% extended 
to solve for rotor acceleration as well.

\subsection{Estimation in Presence of Bounded Data Uncertainties}
\label{BDUdes}
It has been documented in prior literature that network parameter uncertainties are bounded to within $\pm$30\% of their database values \cite{1397479}, while errors in PMU-measured bus frequencies in the U.S.
% bus frequency measurements obtained from PMUs 
lie within $\pm$ 0.008 Hz \cite{10273140}.
% The network parameter uncertainties are bounded within $\pm$30\% of their database values \cite{1397479} and in the bus frequency measurements obtained from PMU is within $\pm$ 0.008 Hz \cite{10273140}. 
To 
% effectively 
incorporate knowledge of these bounds into our problem formulation, we treat \eqref{final_wg} as a BDU problem and
% the estimation problems of this paper as BDU problems.
% Taking \eqref{final_wg} as an example of a BDU problem, we 
solve for $\hat{\omega}_G$ 
% , the unknown parameter, ${\omega}_G$, can be estimated 
by performing
% solving 
the following $\operatorname{min-max}$ optimization: 
% \rajmargin{What do we mean Taking (6) as an example?}
% bounded data uncertainty framework, described in this section,\\
% \eqref{rot_bus_freq} can be expressed as a linear equation,
% \begin{align}
%    A^{-1} \hat{w}_G  = w_{B,me} \label{le}
% \end{align}
% In \eqref{le} $A^{-1} = \mathbb{A}$, and can be solved by the following optimization problem ($ \mathbb{A} \hat{w}_G = w_{B,me}$),
\begin{align*}
\min_{\omega_G} \max_{\substack{\|\delta \mathbb{A}\|_2 \le \eta \\ \|\delta \omega_{B,me}\|_2 \le \eta_b}}
\big\| (\mathbb{A}+\delta \mathbb{A})(\omega_G - \mathbf{1}) - (\omega_{B,me}+\delta \omega_{B,me}) \big\|_2
\end{align*}
where $\mathbb{A} = C^{-1}$, $\delta \mathbf{\mathbb{A}}$ and $\delta {\omega}_{B,me}$ are unknown perturbations in $\mathbb{A}$ and ${\omega}_{B,me}$, and $\eta$ and $\eta_b$ are known bounds on the perturbations. %\rajmargin{why should $\mathbb{A}=C^{-1}$ exist?}
% , respectively. 
% Also, $\|\delta \mathbb{A}\|_2 \leq \eta$ and $\|\delta {\omega}_{B,me}\|_2 \leq \eta_b$, where $\eta$ and $\eta_b$ are the known bounds on the perturbations.
% where, $[\|\delta \mathbb{A}\|_2 \leq \eta, \|\delta w_{B,me}\|_2 \leq \eta_b]$ \\
The solution, $\hat{\omega}_G$, is constructed as follows \cite{doi:10.1137/S0895479896301674}: 
% \rajmargin{In (7) what are we maximizing over?}
\begin{itemize}
    \item \textbf{Perform} singular value decomposition (SVD) of $\mathbb{A}$ and obtain unitary matrices $U \in \mathbb{R}^{M \times M}$ and $V \in \mathbb{R}^{M \times M}$ which represent rotations in space, and diagonal matrix $S \in \mathbb{R}^{M \times M}$ which scales each coordinate by 
    % a factor $\sigma$, (i.e., the singular value of $\mathbb{A}$).
    the singular value of $\mathbb{A}$.
    % Mathematically, this operation is described by $[U, S, V] = \operatorname{SVD}(\mathbb{A})$. 
    % \rajmargin{Why explain a lot on SVD? Perhaps remove stuff on U V and sigma}
    Define $w_{B,me} = U^\top{\omega}_{B,me}$.
    % \item Introducing Singular Value Decomposition (SVD) of $\mathbb{A}$
    % SVD breaks $\mathbf{A}$ into 3 geometrical transformations, U and V $\in \mathbb{R}^{M \times M}$ representing rotation of the space and S $\in \mathbb{R}^{M \times M}$ is a diagonal matrix which is scaling of each coordinate by a factor $\sigma$ (i.e. singular value of A)
    % \begin{align*}
    %     [U, S, V] = \mathbf{SVD}(\mathbb{A})
    % \end{align*} 
    % Now, $\mathbf{w_{B,me}} = U^T*w_{B,me}$
    % \item Introducing Secular Function\\
    \item \textbf{Introduce} the secular function:
    $\mathbb{G(\psi)} = {w_{B,me}}^\top (S^2 - \eta^2 I)(S^2 + \psi I)^{-2}$, 
    where $\psi$ is a positive solution of
    %\eqref{SecFun} the following equation 
    \begin{align}
        \psi = \frac{\eta\sqrt{ \psi^{2}\|(S^2 + \psi I_M)^{-1} {{w}_{B,me}}\|_2}}{\|S(S^2 + \psi I_M)^{-1} {{w}_{B,me}} \|_2}
        \label{SecFun}
    \end{align}
    where $I_M\in R^{M\times M}$ is the identity matrix.
    % $I \in R^{M \times M}$ denotes an identity matrix
    \item \textbf{Define} 
    %terms $\tau_1$ and $\tau_2$ as follows:
    % \item Now defining the terms $\tau_1$ and $\tau_2$ as, \\
    \begin{align*}
        \tau_1 = \frac{\|S^{-1}{w_{B,me}}\|_2}{\|S^{-2}{w_{B,me}}\|_2} \:\:\: \mathrm{and} \:\: \tau_2 = \frac{\|\mathbb{A}^\top {\omega_{B,me}}\|_2}{\|{\omega_{B,me}}\|_2}
    \end{align*}
    % Accounting for variations in $\eta$ which is practically unknown but we know from \cite{1397479} it lies within $\pm$ 0.3.   
With these definitions, the solution $\hat{\omega}_G$ is given by \cite{doi:10.1137/S0895479896301674}:
%Depending on $\eta$ and $\tau_1$ and $\tau_2$, the following options arise:
    \begin{enumerate}
        \item For \( \eta \geq \tau_2 \), the unique solution is \(\hat{\omega}_G  = \mathbf{1}\).
        \item For \( \tau_1 < \eta < \tau_2 \),  the unique solution is given by $\hat{\omega}_G  = (\mathbb{A}^\top \mathbb{A} + \hat{\psi} I_M)^{-1} \mathbb{A}^\top {\omega_{B,me}} + \mathbf{1}$, where $\hat{\psi}$ is the positive root of the secular equation, $\mathbb{G}(\hat{\psi})$ = 0.
        \item For \( \eta \leq \tau_1 \),  the unique solution is \(\hat{\omega}_G = V S^{-1} w_{B,me} +\mathbf{1} = \mathbb{A}^{+}{\omega_{B,me}} + \mathbf{1}\).
        \item For \( \eta = \tau_1 = \tau_2 \),  there are infinitely many solutions: \(\hat{\omega}_G = \beta V S^{-1} w_{B,me} +\mathbf{1} = \beta \mathbb{A}^{+}{\omega_{B,me}} + \mathbf{1}\), for any \( 0\leq \beta \leq 1\).
    \end{enumerate}
\end{itemize} 
% \smallskip 
% After calculating $\hat{\omega}_G$ using bounded data uncertainty, we estimate Inertia, Damping Constant and Mechanical Power in the next section.

The above-mentioned steps are followed for solving all the BDU problems encountered in this paper.

% \section{Inertia, Damping, and Mechanical Power Estimation}
\section{Proposed Approach for Inertia, Damping, and Mechanical Power Input Estimation}
\label{PropAppr}
% This section introduces a simple and efficient least-squares estimation framework to estimate inertia, damping, and mechanical power based on an aggregated continuous-time swing equation. We setup the framework by assuming that speeds and rotor angles of generators are known. In our simulations, we relax this assumption and use the speed and rotor estimates obtained in Section \ref{sec 2}. 

%$\hat{w}_G \approx w_G$ is then substituted in the following equation
% Consider the system of swing equations (expressed compactly in a matrix-vector format) at M network POI buses:  
The swing equation for the  $j$-{th} generator is:
%can be written as:
\begin{align}
2H_j\dot{\omega}_{G,j}(t)+D_j{\omega_{G,j}(t)-p_{m,j}(t)=-p_{e,j}(t)} 
\label{swing_eq}
\end{align}
% \begin{align}
% 2H\dot{w_G}(t)+D{w_G}(t)=p_{m}(t)-p_{\text{e}}(t) \label{swing_eq}
% \end{align}
where $H_j$ denotes its inertia constant, $D_j$ denotes its damping constant, $\dot{\omega}_{G,j}
% \in \mathbb{R}^{T}
$ denotes its rotor acceleration, $\omega_{G,j}$ denotes its rotor speed,
% $w \in \mathbb{R}^{T}$ is rotor speed, 
$p_{m,j}$ denotes its mechanical power input, and $p_{e,j} 
% \in \mathbb{R}^{T}
$ denotes its electrical power output. 
% \rajmargin{$\omega$ is not a vector. The estimation framework is per machine not for all of them! Everything in the swing equation should be scalar.}
%{\in \mathbb{R}^ {1 \times 1}$
% \rajmargin{I thought we are doing inertia estimation per bus. If so, then change the dimensions of vectors and matrices after Eq (11) accordingly}
% \rajmargin{Pm bar and Pe bar are undefined as well}
% In the inertia estimation literature, a common assumption is that $P_m(t)$ is constant. Letting $\bar{M}=2H$ and $p_m(t)=p_m$, for all $t=0,\ldots,T$, we obtain the system of linear equations: 
% We now
Dropping the subscript $j$ and focusing on a single generator for $t=0,\ldots,N$, we get:
% Hereafter, we 
% drop the subscript $j$ and focus on
% % acknowledging that the discussions pertain to 
% a single generator. 
%\rajmargin{Going forward seems redundant!}
% Therefore, by writing \eqref{swing_eq} for $t=0,\ldots,T$, we obtain the following system of linear equations:
% Now, had $p_m$ been a constant, we could write \eqref{swing_eq} for $t=0,\ldots,N$, 
% % for a sufficiently large $N$,
% % where $T$ is the duration of an economic dispatch period, 
% and obtain the following system of linear equations:
\begin{align}\label{eq: linear model}
    \underbrace{\begin{bmatrix}
        \dot{\omega_G}(0) & \omega_G(0) & -1\\
        \dot{\omega_G}(1) & \omega_G(1) & -1\\
        \vdots & \vdots & \vdots \\
        \dot{\omega_G}(N) & \omega_G(N) & -1
    \end{bmatrix}}_{A}\underbrace{\begin{bmatrix}
        \bar{M}\\
        D\\
        \bar{p}_m
    \end{bmatrix}}_{x}&=\underbrace{\begin{bmatrix}
        \bar{p}_e(0)\\
        \bar{p}_e(1)\\
        \vdots \\
        \bar{p}_e(N)
    \end{bmatrix}}_{b} 
\end{align}
% \rajmargin{After the sentence "Now, had $p_m$ has been constant", the negation statement something like "But that is not happening." is missing or not clear}
where $\bar{M}=2H$, $\bar{p}_m=p_m$, and $\bar{p}_e(t)=-p_e(t)$. 
% The three unknown parameters in $x$ can now be computed using the standard least squares estimate for $T > 3$.
It is clear that 
% the three unknown parameters in $x$ can be estimated by solving \eqref{eq: linear model}.
\eqref{eq: linear model} will give reasonable estimates for the three unknown parameters in $x$ only when $p_m$ is a constant.
% as long as 
% $T \geq 3$
% the relation $\bar{p}_m=p_m$ holds.
We now leverage our knowledge of power system operations to better understand 
% the behavior of $p_m$.
the conditions under which $p_m$ varies.
% conditions when $p_m$ remains constant. 

\subsection{Behavior of Mechanical Power Input}
\label{PMechBeh}
\begin{figure}[ht]
\vspace{-0.2cm}
    \centering
    \includegraphics[width=0.48\textwidth]{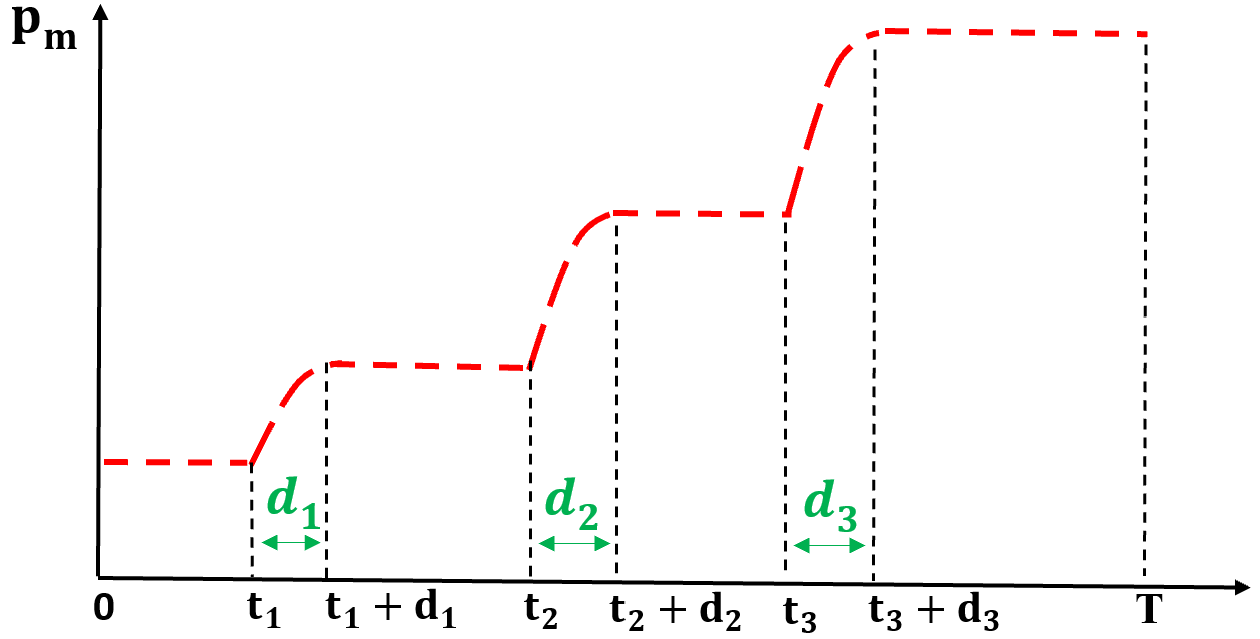}
    \vspace{-0.5em}
    \caption{Realistic variations in mechanical power input}
    \label{fig:p_m_interval}
\end{figure}

The mechanical power input of a generator usually changes in response to the power system's economic dispatch. 
Furthermore, even if the economic dispatch commands coming from the independent system operator have a longer interval (say, $5$ minutes), a utility may want to change their generation at shorter intervals (say, $\mathrm{tens}$ of seconds), 
% to minimize stress on the grid assets.
% This can be the case 
particularly if the system has a large number of CIGs whose inputs are a function of weather conditions.
% This has always been true for synchronous generation, and is now being considered for CIG as their numbers grow. For CIG, this can be done by \textit{not} operating them at maximum-power-point, which leaves them with some head-room/flexibility to participate in dispatch. 
Thus, a typical variation in the $p_m$ of a generator over an economic dispatch period ($T$)
% when responding to dispatch commands 
may look similar to the plot shown in Fig. \ref{fig:p_m_interval}. 
In the figure, $p_m$ ramps considerably in the intervals $d_1$, $d_2$, and $d_3$, and is relatively flat in other time intervals. However, even though it may be flat, its value is not the same.
% it is not the case that the value will be the same. 
Additionally, one cannot measure $p_m$ in real-time using
% there is no way to measure $p_m$ in real-time using an electrical sensor such as 
a PMU. Therefore, even if $p_m$ remains constant, one cannot know its value by direct measurement.

% However, the assumption of constant $p_m(t)$ 
% could be unrealistic as exemplified in Fig.~\ref{fig:p_m_interval}. From this figure, it is clear that $p_m(t)$ is not necessarily constant but is constant for some time intervals and could ramp up and down in other intervals. The ramping behavior is a response to changing load.
 %\rajmargin{replace the y-label $P_m$ with $p_m$, do not overuse the red colors for figures! remove $t_T$}

It is also clear from Fig. \ref{fig:p_m_interval} that $p_m$ can change significantly between the start and end of an economic dispatch period. 
% Furthermore, even if the economic dispatch commands coming from the independent system operator have a longer interval (say, $5$ minutes), a utility may want to operate their generation at shorter intervals (say, $\mathrm{tens}$ of seconds) to minimize stress on the grid assets. 
Lastly, because $p_m$ can change within the economic dispatch period,
% This implies that 
the durations for which $p_m$ is constant (relatively flat in Fig. \ref{fig:p_m_interval}) might not be known in advance.
% These practical considerations not only explain why \eqref{eq: linear model} is unsuitable for inertia estimation when $p_m$ changes, but also give insights on how to circumvent the problem of time-varying $p_m$. 
These practical considerations explain why \eqref{eq: linear model} is unsuitable for inertia estimation when $p_m$ changes. A strategy to circumvent this problem is presented below.

% Owing to the foregoing discussion it is clear that for time-varying mechanical powers $\bar{p}_m(t)$, the linear estimation based on \eqref{eq: linear model} results in bad estimates (see Fig.~\ref{fig:H_wrt_t} and \ref{fig:D_wrt_t}). However, not everything is lost. A simple but an important modification to \eqref{eq: linear model} will improve the estimation accuracy. 

% {\color{blue}Rather than assuming constant mechanical power 
% $\bar{p}_m$ keeps changing at different intervals as shown in Fig. \eqref{fig:p_m_interval} and that causes sudden spikes in the H and D estimation as seen in Figure. \eqref{fig:H_wrt_t} and \eqref{fig:D_wrt_t}. This spike is not because there's some instability in the system it is just because of slight changes in the mechanical power. 
% Now, we consider the samples where $P_m$ is constant. For instance, in Fig. \eqref{fig:p_m_interval} $P_m$ is constant from $t_{k,0}$ to $t_{k,1}$ then it changes at $t_{k,1}$ and remains constant from $t_{k,0}$ to $t_{k,1}$.
% We now measure H where $P_m$ is constant, which can be represented by:}\rajmargin{Paragraph in blue seems incorrect to me.}

\subsection{A Partitioned Formulation for Inertia Estimation}

Since PMUs produce outputs at a much faster rate than the frequency with which $p_m$ changes,
% Therefore, 
one can create rolling windows of duration much smaller than the economic dispatch period ($T$ in Fig. \ref{fig:p_m_interval}), and solve \eqref{eq: linear model} for each window.
% split the economic dispatch period ($T$ in Fig. \ref{fig:p_m_interval}) into smaller time intervals, 
% and solve \eqref{eq: linear model} on a rolling window basis. 
The outcome will be that the estimates over consecutive windows will become \textit{inconsistent} whenever $d_1$, $d_2$, $d_3$ appear inside the windows.
However, the inconsistency of the estimates will reveal the durations over which $p_m$ is varying. 
This information is leveraged below to create a partitioned formulation of the swing equation for inertia estimation. 
% as described below.  

% Therefore, let us partition
% % Partition 
% the discrete-time interval \( t = 0, \ldots, T \) into \( k \) non-overlapping intervals, denoted as \( [t_{i-1}, t_i) \) for \( i = 1, \ldots, k \), where \( 0=t_0, t_1, \ldots, t_{k-1} = T \) are the boundary time points. 
Using the knowledge of when the estimates obtained from \eqref{eq: linear model} have become inconsistent within the economic dispatch period, we partition the discrete-time interval \( t = 0, \ldots, T \) into \( k \) \textit{non-overlapping intervals}, denoted as \( [t_{i-1}+d_{i-1}, t_i) \) for \( i = 1, \ldots, k \), where $t_0=0$, $d_0=0$, and $t_{k} = T$. 
% are the boundary time points.
Note that each of these intervals can have an arbitrary length. 
% \( \Delta t_i
% % = t_i - t_{i-1}
% >0\). 
Now, since $p_m$ remains constant within each interval but may differ between intervals, we have $\bar{p}_m(t)=\bar{p}_{m_i}$ for $t \in [t_{i-1}+d_{i-1}, t_i)$. Finally, for $i \in \{1,\ldots, k\}$, it follows that:
% We assume that the mechanical power, \( \bar{p}_m(t) \), remains constant within each interval \( [t_{i-1}, t_i) \), but may differ between intervals (see Fig.~\ref{fig:p_m_interval}). Thus $\bar{p}_m(t)=\bar{p}_{m_i}$ for $t \in [t_{i-1}, t_i)$. And for $i \in \{1,\ldots, k\}$. it follows that
\begin{align} \label{diff_interval}
    \begin{bmatrix}
        \bar{p}_m(t_0 + d_0)\\
        \vdots \\
        \bar{p}_m(t_1)\\
        \hline
        \bar{p}_m(t_{1} + d_1)\\
        \vdots \\
        \bar{p}_m(t_{2})\\
        \hline
        \vdots \\
        \hline
        \bar{p}_m(t_{k-1} + d_{k-1})\\
        \vdots \\
        \bar{p}_m(t_k)\\
    \end{bmatrix} &=  \underbrace{\begin{bmatrix}
        1 &0 &0 &\hdots &0\\
        1 &0 &0 &\hdots &0\\
        \vdots\\
        1 &0 &0 &\hdots &0\\
        \hline
        0 &1 &0 &\hdots &0\\
        0 &1 &0 &\hdots &0\\
        \vdots\\
        0 &1 &0 &\hdots &0\\
        \hline
        \vdots &\vdots &\vdots &\vdots\\
        \hline
        0 &0 &0 &\hdots &1\\
        0 &0 &0 &\hdots &1\\
        \vdots\\
        0 &0 &0 &\hdots &1\\
    \end{bmatrix}}_{A_2}
    \begin{bmatrix}
        \bar{p}_{m_{1}}\\
        \bar{p}_{m_{2}}\\
        \vdots \\
        \bar{p}_{m_{k}}
    \end{bmatrix}
\end{align}
% \begin{align} \label{diff_interval}
%     \begin{bmatrix}
%         \bar{p}_m(t_{0} + d_1)\\
%         \vdots \\
%         \bar{p}_m(t_{1} + d_1 -1)\\
%         \hline
%         \bar{p}_m(t_{1} + d_2)\\
%         \vdots \\
%         \bar{p}_m(t_{2} + d_2 -1)\\
%         \hline
%         \vdots \\
%         \hline
%         \bar{p}_m(t_{k-1} + d_t)\\
%         \vdots \\
%         \bar{p}_m(t_{k} + d_t -1)\\
%     \end{bmatrix} &=  \underbrace{\begin{bmatrix}
%         -1 &0 &0 &\hdots &0\\
%         -1 &0 &0 &\hdots &0\\
%         \vdots\\
%         -1 &0 &0 &\hdots &0\\
%         \hline
%         0 &-1 &0 &\hdots &0\\
%         0 &-1 &0 &\hdots &0\\
%         \vdots\\
%         0 &-1 &0 &\hdots &0\\
%         \hline
%         \vdots &\vdots &\vdots &\vdots\\
%         \hline
%         0 &0 &0 &\hdots &-1\\
%         0 &0 &0 &\hdots &-1\\
%         \vdots\\
%         0 &0 &0 &\hdots &-1\\
%     \end{bmatrix}}_{A_2}
%     \begin{bmatrix}
%         \bar{p}_{m_{1}}\\
%         \bar{p}_{m_{2}}\\
%         \vdots \\
%         \bar{p}_{m_{k}}
%     \end{bmatrix}
% \end{align}

% Define the vector $\boldsymbol{\omega}_G(t_i)=[\omega_G(t_i),\ldots,\omega_G(t_i-1)]^\top$ and for $i \in \{1,\ldots, k\}$. Similarly, define $\bar{\boldsymbol{p}}_e(t_i)$. Then, using the swing equation in \eqref{swing_eq}, we obtain the linear model 
%Combining \eqref{diff_interval} and \eqref{swing_eq} the formulation is given by,
Based on the partitioned model in \eqref{diff_interval}, we define $\boldsymbol{\omega}_G(t_i)=[\omega_G(t_{i-1}+d_{i-1}),\ldots,\omega_G(t_i)]^\top$ for $i \in \{1,\ldots, k\}$. Similarly, we also define $\dot{\boldsymbol{\omega}}_G(t_i)$ and $\bar{\boldsymbol{p}}_e(t_i)$. Combining these definitions and \eqref{diff_interval} with \eqref{eq: linear model},
% in which $N$ is replaced by $T$, 
we get:
\begin{align}\label{eq: linear model1}
    \underbrace{\begin{bmatrix}
        \dot{\boldsymbol{\omega}}_G(t_1) & \boldsymbol{\omega}_G(t_1) \\
        \dot{\boldsymbol{\omega}}_G(t_2) & \boldsymbol{\omega}_G(t_1) \\
        \vdots & \vdots  \\
        \dot{\boldsymbol{\omega}}_G(t_k) & \boldsymbol{\omega}_G(t_k)
    \end{bmatrix}}_{A_1}\begin{bmatrix}
        \bar{M}\\
        D\\
    \end{bmatrix} - A_2 \begin{bmatrix}
        \bar{p}_{m_{1}}\\
        \bar{p}_{m_{2}}\\
        \vdots \\
        \bar{p}_{m_{k}}
    \end{bmatrix}&=\underbrace{\begin{bmatrix}
        \bar{\boldsymbol{p}}_e(t_1)\\
        \bar{\boldsymbol{p}}_e(t_2)\\
        \vdots \\
        \bar{\boldsymbol{p}}_e(t_k)
    \end{bmatrix}}_{b_1} 
\end{align}

% Note that the 
% % The 
% columns in $A_1$ in \eqref{eq: linear model1} and the first two columns of $A$ in \eqref{eq: linear model} are the same, except that in $A_1$
% % the former 
% we grouped the elements of a column based on partitioned intervals. 
Finally, by rearranging the terms in \eqref{eq: linear model1}, we get: 
\begin{align}\label{eq: final partitioned model}
\begin{bmatrix}
    A_1 & \vert & -A_2
\end{bmatrix}
\begin{bmatrix}
    \bar{M} \\
    D \\
        \hline
    \bar{p}_{m_1} \\
    \vdots \\
    \bar{p}_{m_k}
\end{bmatrix}=b_1
\end{align}
where the number of unknown parameters are $k+2$. 
Therefore, as long as the number of rows of $A_1$, $A_2$, and $b_1$ are greater than $k+2$ (which is always the case since PMUs produce outputs faster  than the frequency with which $p_m$ changes), one can uniquely estimate these parameters.

% In 
% % the partitioned linear model shown in
% \eqref{eq: final partitioned model}, we have $k+2$ unknown parameters. 
% Thus to these parameters, we need the total time $t_k=T>k+2$, which always holds because the partitioned interval length $\Delta T_i$ is at least eight time instants.\footnote{This information on time instants was notified to us by a large scale electric utility company in the USA.} 
% Therefore, as long as the number of rows of $A_1$, $A_2$ and $b$ are greater than $k+2$ (which is always the case since PMUs produce outputs at a much faster rate than the frequency with which $p_m$ changes), one can uniquely estimate these parameters.

Now, since $A_1$ is derived using the methodology described in Section \ref{FreqDC} and $b_1$ is obtained from PMU measurements, $A_1$ and $b_1$ can have bounded uncertainties. Therefore, \eqref{eq: final partitioned model} will be treated as a BDU problem \cite{doi:10.1137/S0895479896301674}  and solved using the methodology described in Section \ref{BDUdes}. Finally, the above-mentioned procedure will be repeated for all the generators present in the system to calculate the system-level inertia constant, $H_{sys}$,
% inertia.
% The equation for calculating system-level inertia is:
using the following relation:
\begin{align}
    \hat{H}_{sys} = \frac{\sum_{j} \hat{H}_{j} S_{j}}{\sum_j S_j}
\end{align}
where, $S_j$ denotes the rated capacity of generator $j$, and $\hat{H}_j$ is the estimated inertia constant of generator $j$ obtained by solving \eqref{eq: final partitioned model}.
Note that the proposed approach estimates system-level inertia constant for the \textit{immediately prior} economic dispatch period. 
This is not a problem during actual implementation because inertia changes slowly and a delay of one economic dispatch period for getting accurate and consistent inertia estimates is an acceptable compromise.

% \eqref{eq: linear model} is used to estimate H, D and $P_m$ for the intervals where $P_m$ is stable.

% \newpage 
\section{Simulation Results}
We evaluate the proposed approach on the IEEE 14-bus and 39-bus systems.
The former has no CIGs, but the latter has three CIGs at buses 2, 29, and 39.
% , showing the universal applicability of the proposed approach.
% A few of our  buses are modeled as renewable energy sources.
We generate data by creating random fluctuations in the loads by varying them within $\pm0.1\%$ of their actual value \cite{10026331}. 
% simulating the morning load pick-up during which loads increase by  approximately 60\% over one hour \cite{6184586}.
Bounded perturbations specified in \cite{1397479} and \cite{10273140} are added to the network parameters and PMU measurements, respectively.
The $p_e$ of the generators is changed every $8$ seconds, and the dynamic simulations are run for $40$ seconds in PSS/E.
We use the absolute relative error ($\mathrm{ARE}$) metric 
% as the evaluation criterion. 
for evaluating the performance.%\rajmargin{In the second sentence from the last in the first paragraph of IV, do we mean to say $p_m$?}
%is used to evaluate the performance of the proposed approach.
% Simulation data was generated by based on insights derived from morning load pickup patterns, during which peak demand occurs. Observations indicate that load increases by approximately 60 \% within an hour during this period. Power utilities typically adjust load levels at intervals of at least 8 seconds, allowing for load changes at each interval. These load changes can be modeled as a Gaussian distribution with a standard deviation of 0.0004.

% For these simulations, load variations are modeled using a Gaussian distribution with a standard deviation of 1\%, representing fluctuations over an hourly timeframe. 
% Load changes are applied at intervals reflective of typical utility practices, specifically every 8 seconds, to mimic real-world scenarios where power utilities adjust loads.
\vspace{-0.09cm}
\subsection{IEEE 14-Bus System}\label{14b}
\begin{figure}[ht]
    \centering
\includegraphics[width=0.485\textwidth]{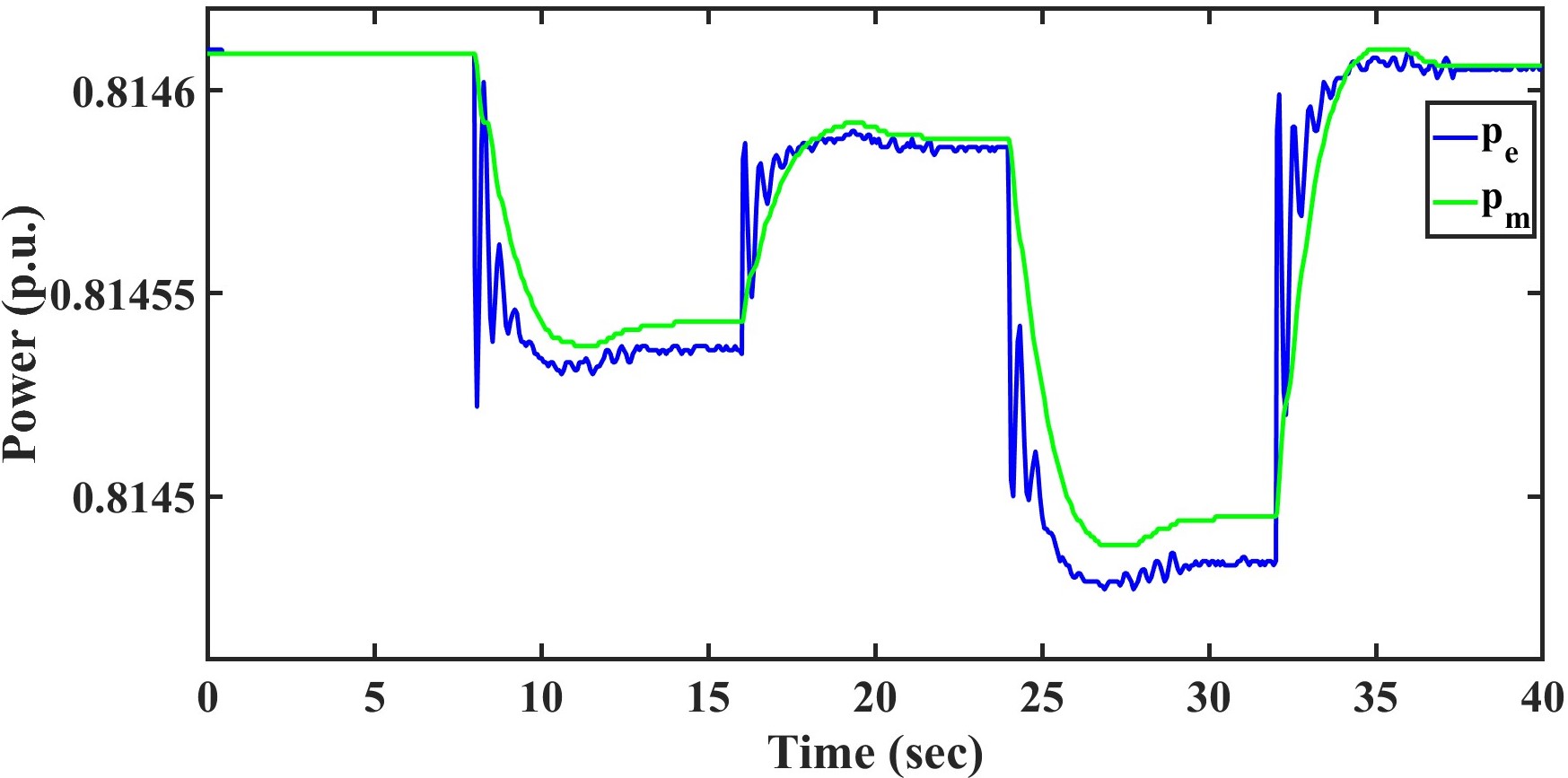}
    \vspace{-1.5em}
    \caption{$p_{e}$ vs. $p_{m}$ (in p.u.) on the same base for $G_1$ }
    \label{fig:pepm14}
\end{figure}

\begin{figure}[ht]
    \centering
\includegraphics[width=0.485\textwidth]{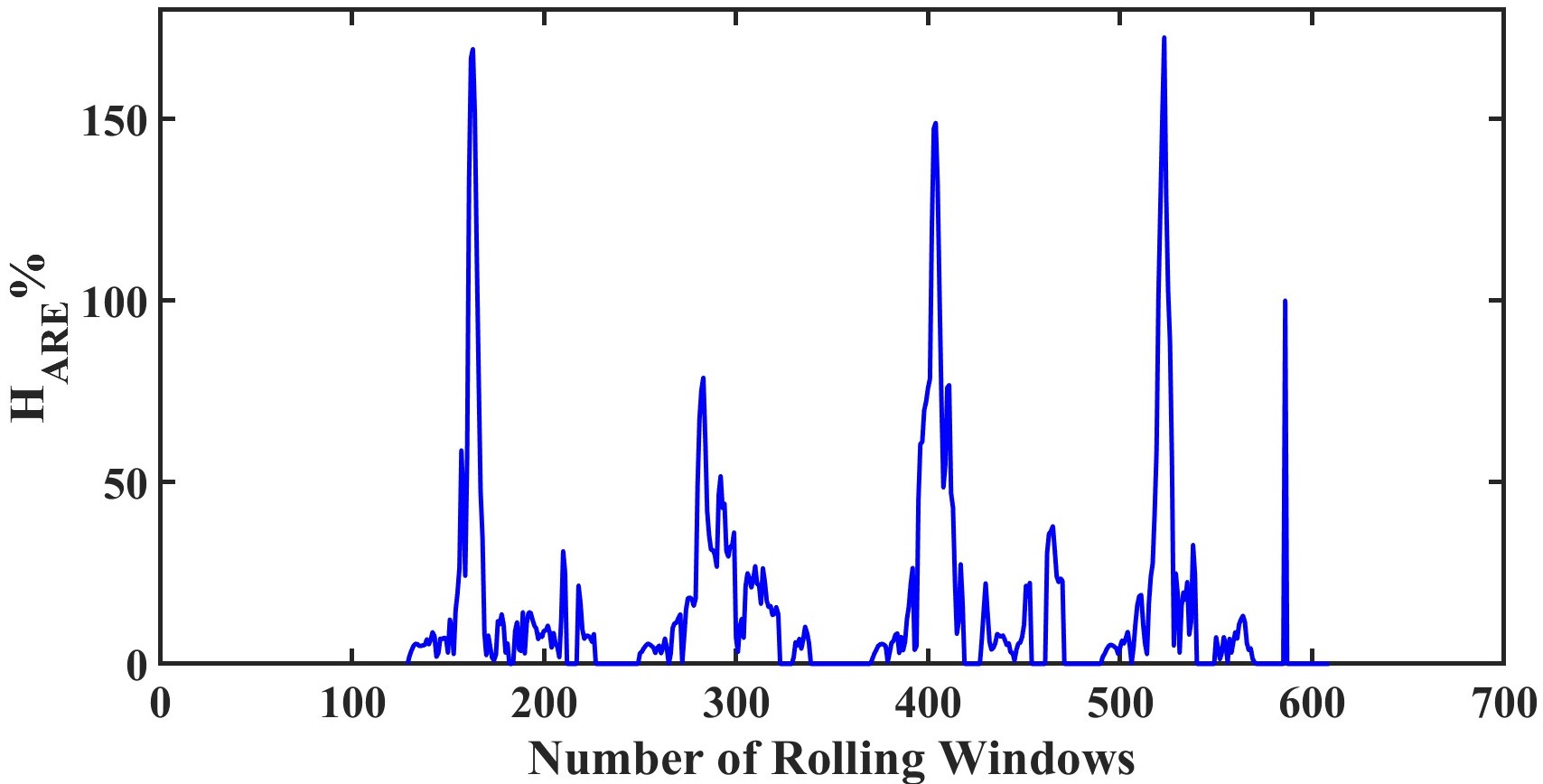}
    \vspace{-1.5em}    \caption{$H_{\mathrm{ARE}}$ (in $\%$) for $G_1$ in presence of variable mechanical power input}
    \label{fig:hvst14_m}
    \vspace{-0.5cm}
\end{figure}

\begin{figure}[ht]
    \centering
    \includegraphics[width=0.485\textwidth]{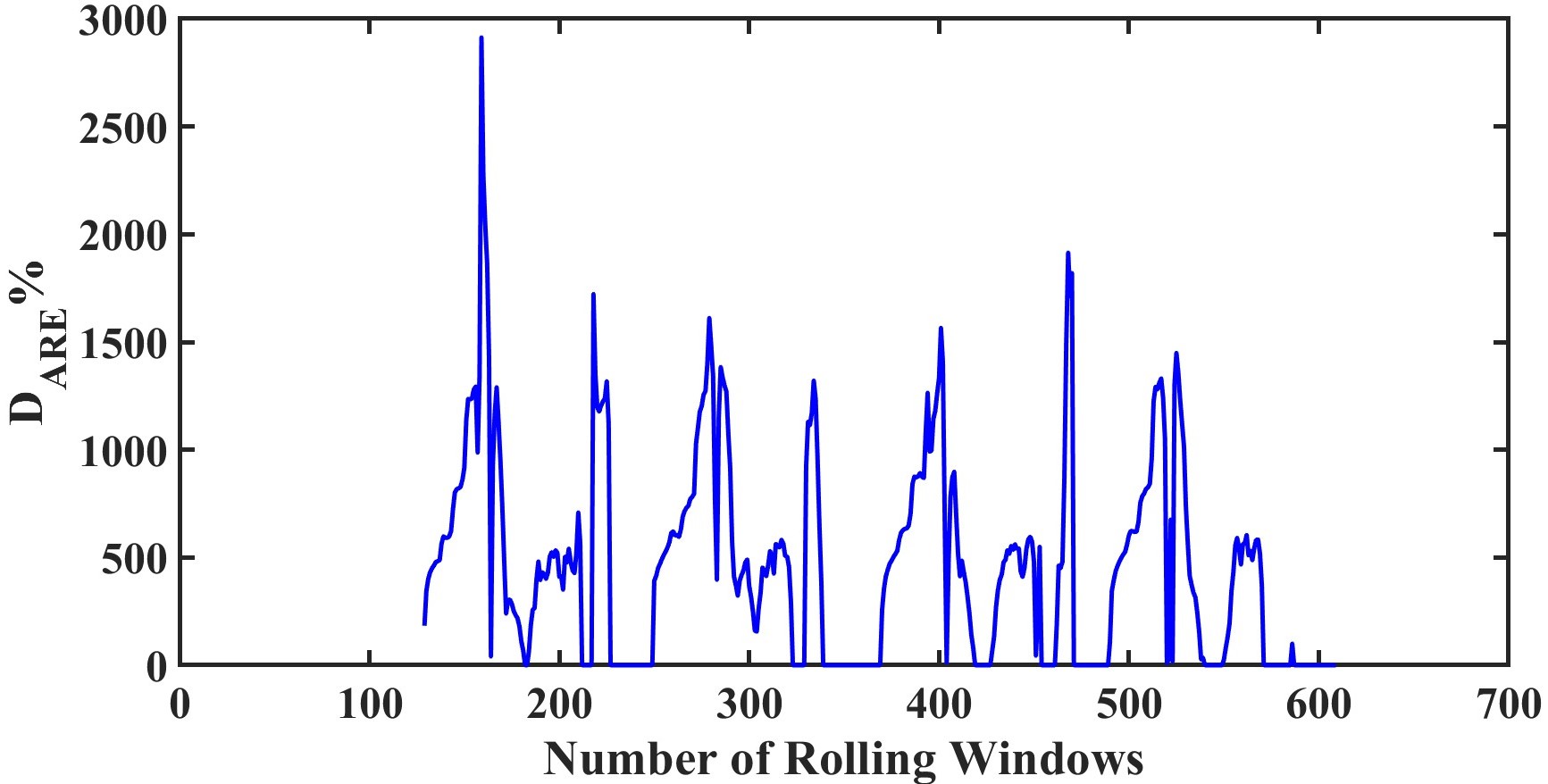}
    \vspace{-1.5em}
    \caption{$D_{\mathrm{ARE}}$ (in $\%$) for $G_1$ for variable mechanical power input}
    \label{fig:Dvst14_m}
\end{figure}

% The IEEE 14-bus system, 
% which includes 
This system has five generators labeled \{$G_1$, $G_2$, $G_3$, $G_4$, $G_5$\}. 
% connected to buses \{1, 2, 3, 6, 8\},
% and supplying loads at buses \{2, 3, 4, 5, 6, 9, 10, 11, 12, 13, 14\}, 
%serves as the first test case to validate the proposed approach. 
The variation in $p_e$ and $p_m$ for $G_1$
% for a change every $8$ seconds over a $40$-second period 
is shown in Fig. \ref{fig:pepm14}.
% To estimate 
% % the inertia constant 
% $H$ and 
% % damping constant 
% $D$, 
However, the knowledge of the interval after which the change occurs ($8$ seconds, in this case), is not known a priori.
Therefore, we first solve \eqref{eq: linear model} over the entire $40$-second time interval using a rolling window of size $0.33$ seconds. Figs. \ref{fig:hvst14_m} and \ref{fig:Dvst14_m} demonstrate that the fluctuations in $p_m$ (shown in Fig. \ref{fig:pepm14}) 
% (shown in Fig. \ref{fig:pvst14_m}) 
introduce extremely high variability in the estimates of $H$ and $D$.
Note that the regions of the plots of $\%H_{\mathrm{ARE}}$ and $\%D_{\mathrm{ARE}}$ which are blank (e.g., first $120$ rolling windows in Fig. \ref{fig:Dvst14_m}) were due to the estimation algorithm not converging because of the ill-conditioning of the linear system of equations.
% The ill-conditioning occurs if the power system is completely stationary-in-time, which is unlikely to happen in practice.
One way to avoid this problem 
% altogether (even in simulations) 
is by increasing the size of the rolling window.
The optimal window size should be such that (i) the linear estimation problems derived in Section \ref{PropAppr} are rank-sufficient, and (ii) there are multiple windows in which $p_m$ has the time to settle down and become flat.

% , with notable spikes occurring due to changes in $P_m$.
%\begin{figure}[ht!]
 %   \centering
 %   \includegraphics[width=0.5\textwidth]{p14bus_time.jpg}
%    \vspace{-2mm}
%    \caption{$P_{mech}$ at Bus 5}
 %   \label{fig:p14_m}
%\end{figure}

% \begin{figure}
%     \centering
% \includegraphics[width=0.485\textwidth]{pm_b1_14.jpg}
%     \vspace{-1.5em}
%     \caption{Comparison of $p_m$ and $\hat{p}_m$ (in p.u.) for $G_3$}
%     \label{fig:pvst14_m}
% \end{figure}

Now, by analyzing the variability in the estimates of $H$ and $D$, we identify time intervals where $p_m$ remains constant. Then, we use \eqref{eq: final partitioned model} to estimate 
% $H$ and $D$ across the 40-second time interval, and $p_m$ for those time intervals in which it is a constant.
$H$, $D$, and $p_m$ for those time intervals 
% in which $p_m$ is a constant.
only.
% analyze time intervals where $p_m$ remains constant, and apply \eqref{eq: linear model} across various segments where $P_m$ stability is maintained. The absolute relative error (ARE) calculated by,
% \begin{align}
% H_{\mathrm{ARE}} = \left|\frac{\hat{H} - H_{true}}{H_{true}} \right|
% \end{align}
% where, $\hat{H}$ represents the estimated value of $H$,
% ARE of $H$ and $D$ across the stable $p_m$ intervals, presented in Table \ref{tab14bus}, indicate that $H$ and $D$ remain consistent under stable conditions of $P_m$ and any change in the system due to changing $P_m$ can be reflected by $H$ and $D$ estimates. $P_{m, avg}$ denotes the average ARE of all the intervals where $P_m$ remains constant.
%However, as system instabilities emerge, the values of $H$ and $D$ reflect these changes, underscoring the method’s sensitivity to system dynamics and its capacity to capture transient instabilities.
The results are shown in Table \ref{tab14bus} for all five generators of the 14-bus system. Note that the entries in the last column indicate the average $\%\mathrm{ARE}$ for $p_m$ across different time intervals.
The very low values of the $\%\mathrm{ARE}$ confirm the robustness of the proposed approach.

%(less than $10^{-3}$)
\begin{table}
\vspace{-0.5cm}
\caption{IEEE 14-bus system: $\mathrm{ARE}$ (in $\%$) for $H$, $D$, and $p_m$ using partitioned form of swing equation}
\vspace{-1.5em}
\begin{center}
\begin{tabular}{|c|c|c|c|}
\hline
\multirow{2}{*}{\textbf{Generator}} & \multicolumn{3}{|c|}{$\mathrm{ARE}$ in $\%$} \\
\cline{2-4}
& {H} & {D} & {$p_{m,avg}$}\\
\hline
1 & 0.0115 & 0.41 & 0.6\\
\hline
2 & 0.0322 & 0.24 & 0.3\\
\hline
3 & 0.0198 & 0.21 & 0.09\\
\hline
4 & 0.0272 & 0.17 & 0.4\\
\hline
5 & 0.0121 & 0.20 & 0.5\\
\hline
\end{tabular}
\label{tab14bus}
\end{center}
\end{table}

%\begin{table}[ht!]
%\caption{$P_m$ Estimates in P.U. - 14 Bus System}
%\begin{center}
%\begin{tabular}{|c|c|c|}
%\hline
%\textbf{Generator} & \textbf{TV} & \textbf{Estimated} \\
%\hline
%1 & 0.8146 & 0.8144 \\
%\hline
%2 & 0.4000 & 0.4018 \\
%\hline
%3 & 0.4000 & 0.4002 \\
%\hline
%4 & 0.3000 & 0.3020 \\
%\hline
%5 & 0.3500 & 0.3505 \\
%\hline
%\end{tabular}
%\label{tab14_pm_estimatebus}
%\end{center}
%\end{table}

\subsection{IEEE 39-Bus System with Renewable Energy Resources}
\label{39b}
\begin{figure}[ht]
    \centering
    \includegraphics[width=0.485\textwidth]{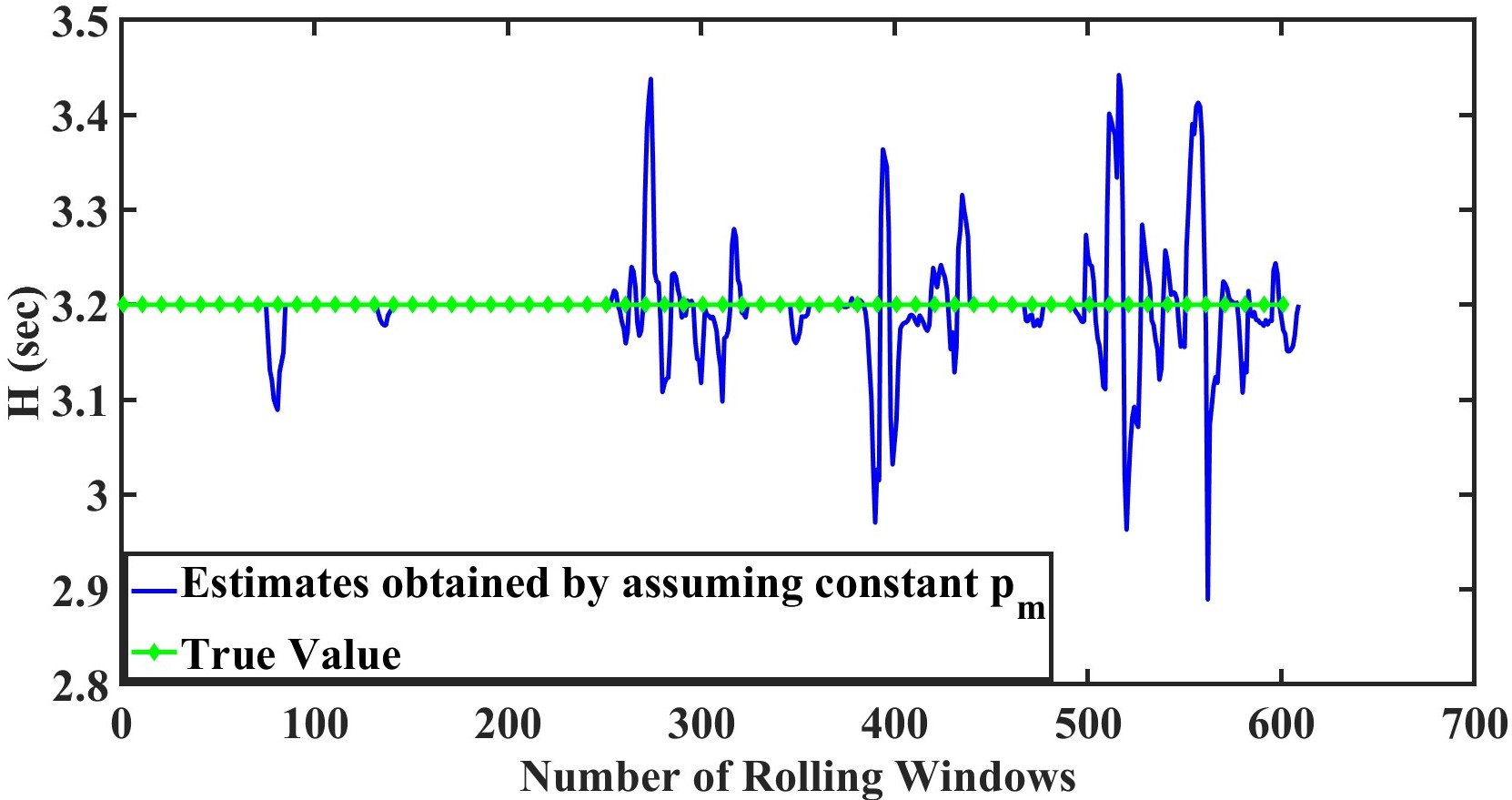}
    \vspace{-1.5em}    \caption{Fluctuations in $\hat{H}$ for $G_3$ for varying mechanical power input}
    \label{fig:H_wrt_t}
    \vspace{-0.5cm}
\end{figure}

\begin{figure}
\vspace{-0.5cm}
    \centering
    \includegraphics[width=0.485\textwidth]{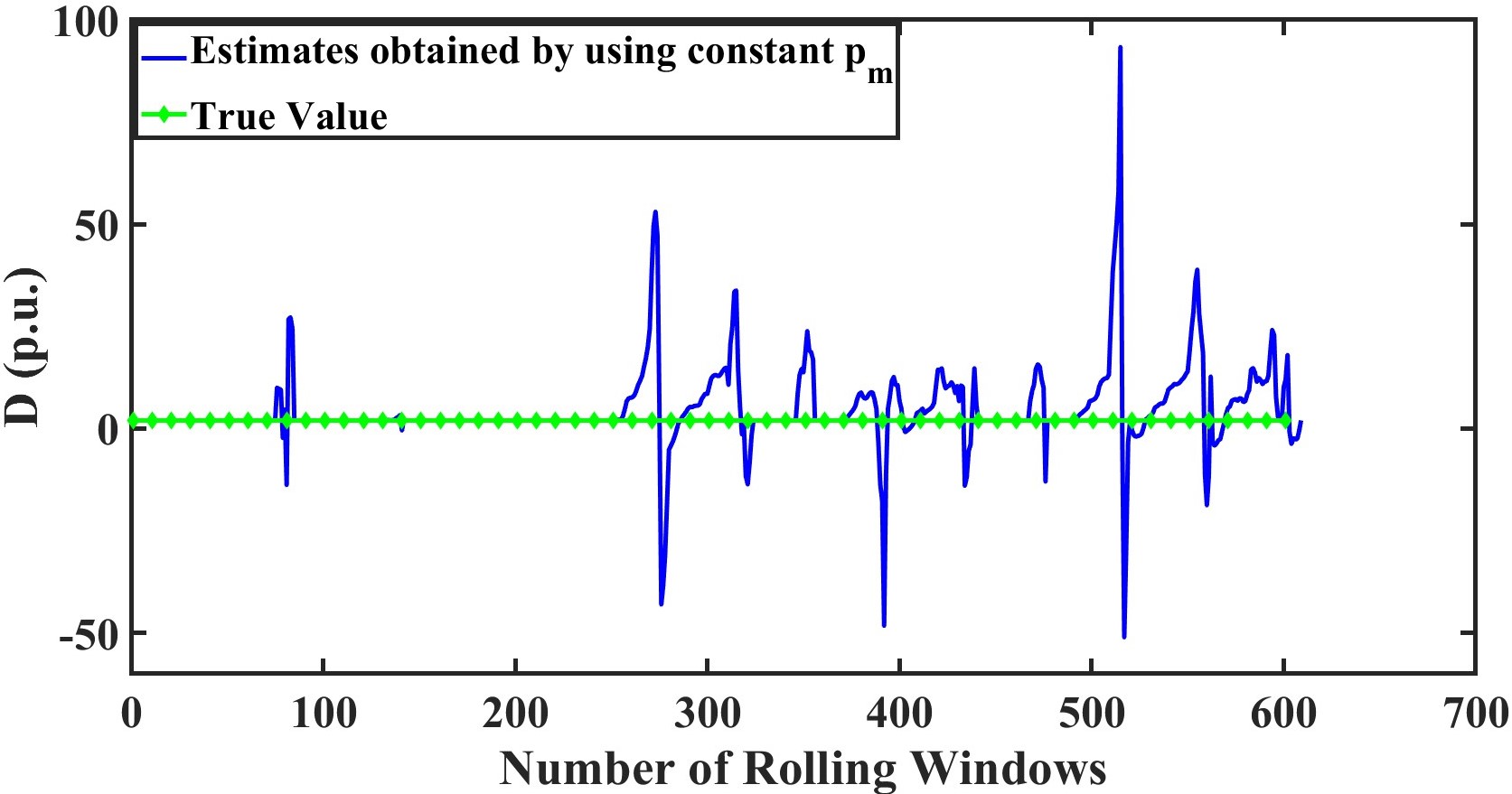}
    \vspace{-2.0em}    \caption{Fluctuations in $D$ estimates for $G_3$ in presence of varying mechanical power input}
    \label{fig:D_wrt_t}
\end{figure}

The modified IEEE 39-bus system
% with renewable energy sources 
consists of 3 renewable resources and
% i.e. wind machines, 
10 generators labeled \{$G_1$, $\hdots$, $G_{10}$\}. 
% connected to buses \{30, $\hdots$ 39\}.
% and 19 loads \{3, 4, 7, 8, 12, 15, 16, 18, 20, 21, 23, 24, 25, 26, 27, 28, 29, 31, 39\}.
Estimation of $H$ and $D$ using \eqref{eq: linear model} over the $40$-second interval results in Figs. \ref{fig:H_wrt_t} and \ref{fig:D_wrt_t}. The figures confirm
% Figures \eqref{fig:H_wrt_t} and \eqref{fig:D_wrt_t} demonstrate 
that the estimated values of $H$ and $D$ fluctuate considerably.
Note that in contrast to Figs. \ref{fig:hvst14_m} and \ref{fig:Dvst14_m} which showed the fluctuations in the \%$\mathrm{ARE}$, Figs. \ref{fig:H_wrt_t} and \ref{fig:D_wrt_t} depict the variations in the actual estimates. However, the inferences drawn are the same (namely, that changes in $p_m$ severely deteriorate estimation performance).
% with changes in $p_m$. 
% (see Fig. \ref{fig:Pm_wrt_t}).
By examining these plots, 
% we identify intervals where $p_m$ remains flat and substitute these intervals in \eqref{eq: final partitioned model}.
intervals in which $p_m$ remains constant are identified and the remaining rolling windows are discarded.
The $\%\mathrm{ARE}$ of $H$, $D$, and $p_m$ estimates obtained using \eqref{eq: final partitioned model}, after omitting the intervals where $p_m$ is varying, are shown in Table \ref{tab39bus}. 
% The accuracy of $H$ and $D$ estimates indicate that when $p_m$ is constant,
% % under stable $p_m$ 
% it is feasible to estimate $H$ and $D$ without the need to assume $p_m$ is constant throughout the estimation interval. 
The $\%\mathrm{ARE}$ values (particularly for $\hat{H}$) are even lower for this system confirming the accuracy and widespread applicability 
% efficiency 
of the proposed approach. %\rajmargin{are figures 6 7 8 estimates or the actual values?}
%can be applied to any type of power system. 

% \begin{figure}
%     \centering
%     \includegraphics[width=0.485\textwidth]{pm_39_bs3.jpg}
%     \vspace{-1.5em}
%     \caption{Variability in $p_m$ estimates using conventional swing equation at $G_3$}
%     \label{fig:Pm_wrt_t}
% \end{figure}

\begin{table}
\caption{IEEE 39-bus system: $\mathrm{ARE}$ (in $\%$) for $H$, $D$, and $p_m$ using partitioned form of swing equation}
\vspace{-1.5em}
\begin{center}
\begin{tabular}{|c|c|c|c|}
\hline
\multirow{2}{*}{\textbf{Generator}} & \multicolumn{3}{|c|}{$\mathrm{ARE}$ in {$\%$}} \\
\cline{2-4}
& {H $\times 10^{-4}$} & {D} & {$p_{m, avg}$}\\
\hline
1 & 3.6592 & 0.0853 & 0.0292\\
\hline
2 & 3.4466 & 1.51 & 0.0105\\
\hline
3 & 3.6608 & 0.15 & 0.0405\\
\hline
4 & 1.1032 & 0.11 & 0.0014\\
\hline
5 & 4.9260 & 0.31 & 0.0441\\
\hline
6 & 2.7518 & 0.26 & 0.0889\\
\hline
7 & 2.1293 & 0.23 & 0.0774\\
\hline
8 & 2.4934 & 0.93 & 0.0358\\
\hline
9 & 4.7090 & 1.66 & 0.0808\\
\hline
10 & 4.2717 & 1.49 & 0.0784\\
\hline
\end{tabular}
\label{tab39bus}
\end{center}
\vspace{-0.5cm}
\end{table}

Lastly, Table \ref{tab:system_inertia} presents the system-level inertia estimates for the IEEE 14-bus and 39-bus systems. 
As seen from the table, the calculated system-level inertia constant values are very close to the true values (obtained from PSS/E dynamic data for the two test systems),
% highly consistent with the true values, 
demonstrating the effectiveness of the proposed inertia estimation method in presence of bounded uncertainties in the network parameters and PMU measurements and changing mechanical power inputs.

\begin{table}
\caption{Estimated system-level inertia constant ($\hat{H}_{sys}$) using partitioned form of swing equation}
\vspace{-1.5em}
\begin{center}
\begin{tabular}{|c|c|c|}
\hline
\textbf{Test System} & {True value of $\hat{H}_{sys}$} & {Estimated value of $\hat{H}_{sys}$} \\ \hline
14-bus           & 4.9400 seconds& 4.9404 seconds\\ \hline
39-bus          & 3.2000 seconds& 3.1999 seconds \\ \hline
\end{tabular}
\label{tab:system_inertia}
\end{center}
\vspace{-0.5cm}
\end{table}
%\begin{table}[ht!]
%\caption{$P_m$ Estimates in P.U. - 39 Bus System}
%\begin{center}
%\begin{tabular}{|c|c|c|}
%\hline
%\textbf{Generator} & \textbf{TV} & \textbf{Estimated} \\
%\hline
%1 & 0.9106 &  0.9107\\
%\hline
%2 & 0.7050 &  0.7053\\
%\hline
%3 & 0.5345 &  0.5344\\
%\hline
%4 & 0.5386 &  0.5388\\
%\hline
%5 & 0.4707 &  0.4708\\
%\hline
%6 & 0.4272 &  0.4274\\
%\hline
%7 & 0.5469 & 5470 \\
%\hline
%8 & 0.5586 &  0.5588\\
%\hline
%9 & 0.3745 &  0.3747\\
%\hline
%10 & 0.5912 &  0.5914\\
%\hline
%\end{tabular}
%\label{tab39bus_pm_estimate}
%\end{center}
%\end{table}

\section{Conclusion}

This paper presents a novel approach for estimating three parameters---namely, inertia constant, damping constant, and mechanical power input---of every generator from ambient PMU data.
The proposed approach relaxes the constant mechanical power input assumption and accounts for the presence of bounded uncertainties in the network parameters and PMU measurements.
First, rotor speed and acceleration are estimated for all the generators from PMUs placed only at the POI buses by embedding frequency spatial correlations into the FDF.
Then, these estimates are inserted into the swing equation to check for consistency of the inertia and damping estimates.
Considerable variations in the resulting estimates are indicators of changes in mechanical power inputs.
This knowledge is used to determine intervals for which the mechanical power input does not change.
Finally, these intervals are fed into a partitioned form of the swing equation to estimate the three desired parameters as well as the system-level inertia constant.
The results obtained using two IEEE test systems affirm the accuracy, robustness, and widespread applicability of the proposed approach.
% In this paper we present a novel approach to estimate inertia, damping and mechanical power. Our method relaxes the constant mechanical power assumption and considers the bounded uncertainties in the PMU measurements and network parameters. The high variability of $H$ and $D$ estimates measured using the swing equation shows that $H$ and $D$ estimates can be used to detect system imbalance and can serve as indicators of system disturbances. We recommend pausing the estimation process whenever these estimates deviate significantly and resuming only after the mechanical power stabilizes to a constant value. In future work, we aim to develop a change-point detection method to identify variations in mechanical power more precisely. This enhancement will allow our approach to be applied effectively to larger and more complex power systems.
The focus of future research will be on investigating this approach in the presence of intermittency and variability typically encountered by CIGs in the field.

%In this paper, we present a novel approach for estimating system inertia, damping constant and mechanical power, with a focus on detecting imbalances in power systems.  When a disturbance in mechanical power is identified, the approach recommends a waiting period before resuming estimation, allowing variations in mechanical power to stabilize. Our approach accommodates fluctuations in mechanical power and incorporates damping constant estimation. Additionally, it leverages bounded data uncertainty, utilizing prior bounds on data to enhance estimation robustness and reliability. In future work, we aim to develop a change-point detection method to identify variations in mechanical power more precisely. This enhancement will allow our approach to be applied effectively to larger and more complex power systems.

%\section*{Acknowledgment}
\vspace{-0.2cm}
\bibliographystyle{IEEEtran}
\bibliography{bibref}

\end{document}